\documentclass[sigconf]{acmart}
\usepackage{color,xcolor}
\usepackage{multirow}
\usepackage{booktabs} 
\usepackage{tablefootnote}
\usepackage{ifthen}
\usepackage{url}
\usepackage{amsmath}
\usepackage{hyphenat}
\usepackage{bbding}
\usepackage{xspace}
\usepackage[T1]{fontenc}
\usepackage[ruled,linesnumbered]{algorithm2e}
\usepackage{array,multirow,graphicx}
\usepackage{float}
\usepackage{balance}
\usepackage{hyperref}
\usepackage{tikz}
\usepackage{calc}
\usepackage{subfigure}
\usepackage{listings,amsfonts}
\usepackage{amsmath,pifont}
\usepackage{url}
\usepackage{xspace}
\usepackage{hyperref,endnotes}
\usepackage{array}
\usepackage{multirow,makecell}
\usepackage{pgfplots}
\usepackage{xcolor}
\usepackage{courier}
\usepackage{hyphenat}

\pgfplotsset{width=7cm,compat=1.16}
\usepgfplotslibrary{statistics}
\graphicspath{ {./} }

\definecolor{dkgreen}{rgb}{0,0.6,0}
\definecolor{gray}{rgb}{0.4,0.4,0.4}
\definecolor{mauve}{rgb}{0.58,0,0.82}
\definecolor{darkblue}{rgb}{0.0,0.0,0.6}
\definecolor{lightblue}{rgb}{0.0,0.0,0.9}
\definecolor{cyan}{rgb}{0.0,0.6,0.6}
\definecolor{darkred}{rgb}{0.6,0.0,0.0}

\definecolor{yellow}{RGB}{255,255,153}
\definecolor{grey}{RGB}{220,220,220}
\definecolor{green}{RGB}{0,100,0}

\definecolor{KWColor}{rgb}{0.37,0.08,0.25}
\definecolor{CommentColor}{rgb}{0.133,0.545,0.133}
\definecolor{StringColor}{rgb}{0,0.126,0.941}
\definecolor{commentgreen}{RGB}{2,112,10}
\definecolor{eminence}{RGB}{108,48,130}
\definecolor{weborange}{RGB}{255,165,0}
\definecolor{frenchplum}{RGB}{129,20,83}

\lstdefinelanguage{XML}
{
  morestring=[s][\color{mauve}]{"}{"},
  morestring=[s][\color{black}]{>}{<},
  morecomment=[s]{<?}{?>},
  morecomment=[s][\color{dkgreen}]{<!--}{-->},
  stringstyle=\color{black},
  identifierstyle=\color{lightblue},
  keywordstyle=\color{red},
  morekeywords={xmlns,xsi,noNamespaceSchemaLocation,type,id,x,y,source,target,version,tool,transRef,roleRef,objective,eventually},
  texcl=true
}

\definecolor{yellow}{RGB}{255,255,153}
\definecolor{grey}{RGB}{224,224,224}
\definecolor{green}{RGB}{0,100,0}

\newboolean{showcomments}
\setboolean{showcomments}{true}
\ifthenelse{\boolean{showcomments}}
 { \newcommand{\mynote}[2]{
      \fbox{\bfseries\sffamily\scriptsize#1}
        {\small$\blacktriangleright$\textsf{\emph{#2}}$\blacktriangleleft$}}}
        { \newcommand{\mynote}[2]{}}

\definecolor{DarkOrange}{rgb}{0.8,0.3,0.0} 
\definecolor{DarkCyan}{rgb}{0.0, 0.55, 0.55}
\definecolor{codegreen}{rgb}{0,0.6,0}
\definecolor{codegray}{rgb}{0.5,0.5,0.5}
\definecolor{codepurple}{rgb}{0.58,0,0.82}
\definecolor{backcolour}{rgb}{0.95,0.95,0.92}

\newcommand\tool[1]{\emph{AndroMevol}}

\copyrightyear{2022}
\acmYear{2022}
\setcopyright{acmcopyright}\acmConference[ASE '22]{37th IEEE/ACM International Conference on Automated Software Engineering}{October 10--14, 2022}{Rochester, MI, USA}
\acmBooktitle{37th IEEE/ACM International Conference on Automated Software Engineering (ASE '22), October 10--14, 2022, Rochester, MI, USA}
\acmPrice{15.00}
\acmDOI{10.1145/3551349.3561341}
\acmISBN{978-1-4503-9475-8/22/10}

\begin{document}

\title{A First Look at CI/CD Adoptions in Open-Source Android Apps}

\author{Pei Liu, Xiaoyu Sun, Yanjie Zhao, Yonghui Liu, John Grundy, and Li Li}
\email{{Pei.Liu,Xiaoyu.Sun,Yanjie.Zhao,Yonghui.Liu,John.Grundy,Li.Li}@monash.edu}
\affiliation{
  \institution{Monash University}
  \country{Australia}}
  \authornote{Corresponding author.}
  




\begin{abstract}
Continuous Integration (CI) and Continuous Delivery (CD) have been demonstrated to be effective in facilitating software building, testing, and deployment. Many research studies have  investigated and subsequently improved their working processes. Unfortunately, such research efforts have largely not touched on the usage of CI/CD in the development of Android apps. We fill this gap by conducting an exploratory study of CI/CD adoption in open-source Android apps.
We start by collecting a set of 84,475 open-source Android apps from the most popular three online code hosting sites, namely Github, GitLab, and Bitbucket. 
We then look into those apps and find that (1) only around 10\% of apps have leveraged CI/CD services, i.e., the majority of open-source Android apps are developed without accessing CI/CD services, (2) a small number of apps (291) has even adopted multiple CI/CD services, (3) nearly half of the apps adopted CI/CD services have not really used them, and (4) CI/CD services are useful to improve the popularity of projects.
\end{abstract}




\maketitle

\section{Introduction}
\label{sec:introduction}

Modern software development involves a variety of platforms, third-party libraries, tools, and a large group of developers, who always need to maintain a consistent development environment to integrate and validate local changes. 
However, it is non-trivial to achieve that, i.e., significant efforts of developers are often required.
To mitigate this, developers has invented  Continuous Integration (CI) and Continuous Delivery (CD) processes.
CI/CD are a set of software development practices that enforce automation in building, testing, and deployment or delivery of applications and help developers to deliver code changes more timely and reliably~\cite{cicddefinition,duvall2007continuous}. 
Specifically, CI aims to facilitate such arduous maintenance with an automated approach to build, package, and test their application updates in a consistent environment. 
CD then takes over once CI finishes to deliver the code into production so as to shorten the deployment cycle. This is greatly demanded as nowadays popular software is released at an unprecedented pace to try to keep an edge against the competition with similar software.
For example, Facebook, over a period of 4 years, has greatly shortened the deployment cycle to once a week from every eight weeks~\cite{rossi2016continuous}. Google chrome also shortened its release cycle to four weeks from six weeks~\cite{googlerelease}.

The convenience of CI/CD has already been leveraged both by open-source~\cite{hilton2016usage} and proprietary software development teams, such as Google, Mozilla, Microsoft, etc. With the help of CI/CD, the HP LaserJet Firmware division announced that it helps them reduce development costs by 78\%~\cite{cdevidence}. The widespread adoption of CI/CD enlightens the emergence of cloud-based CI/CD services, such as \textsc{CircleCI}~\cite{circleci}, \textsc{TravisCI}~\cite{travisci}, and Github Actions, which provide consumers the benefits of CI/CD without the additional efforts to maintain the infrastructures of these online CI/CD services.

Other researchers have also seen the great potential of CI/CD and hence have spent significant efforts in studying and improving its working process.
For example, Zhang et al.~\cite{zhang2022buildsheriff} did a large-scale empirical study on 163,371 test failure builds in real-world Java projects utilizing~\textsc{TravisCI} to triage the failure CI builds. They proposed an approach called \textsc{BuildSheriff} to classify the test failures with the same root cause into the same group. 
Hilton et al.~\cite{hilton2016usage} utilized three complementary methods to investigate the usage of CI in open-source projects. They analyzed 1,529,291 builds generated from the well-known CI provider~\textsc{TravisCI} among 34,544 Github projects, and surveyed 442 developers. Their experimental results concluded that projects with CI are released twice as faster as the ones without. The developers also manage to accept pull requests faster and generally are less worried about build failures.
Gallaba et al.~\cite{gallaba2018noise} conducted an extensive empirical study on 3.7 million build jobs from 1,276 open-source projects adopting~\textsc{TravisCI}. 
They observed that some passing builds contain ignored failures, that certain builds have misleading and erroneous results, and that some failing builds have passing jobs.
Their findings rejected two critical assumptions: build results are not noisy, and builds are equal. In addition, Gallaba et al.~\cite{gallaba2022lessons} also explored the 23.3 million builds spanning 7,795 open-source projects utilizing~\textsc{CircleCI} from 2012 to 2020. 
Their quantitative analysis reveals that approaches to accelerate builds are necessary to boost the adoption of CI/CD services. The robustness of CI services would be enhanced via the optimisation of compilation and testing procedures, the elimination of improper configurations, and the resolution of service availability problems.

Unfortunately, despite the fact that various efforts have been put into studying CI/CD in open-source projects, our community has not yet explored the adoption of CI/CD in Android apps, one of the most popular types of software.
In this paper, we propose to bridge this gap by conducting the first empirical study about CI/CD adoptions in open-source Android apps.
We start by collecting a dataset of open-source Android apps from well-known online repository hosting sites, including Github, GitLab, and Bitbucket.
This step results in 84,475 open-source Android apps, which are then leveraged to fulfill our research objective of understanding the adoption of CI/CD services in open-source Android apps.
Our exploratory study has revealed several interesting findings: 1) The majority (89.94\%) of our collected Android projects do not access any CI/CD services,
2) Concerning the projects configured with CI/CD services, they are generally more popular (e.g., more Github stars) than those projects that do not access CI/CD services,
3) Among the apps that adopted CI/CD services, only 59.58\% of them have their CI/CD processes executed practically during the development,
4) The apps with practical usage of CI/CD are even more popular than those apps that only adopt CI/CD but do not configure it properly to allow successful execution of the CI/CD processes.

Our work makes the following main contributions.
\begin{itemize}
    \item We release to the community a set of 84,475 open-source Android apps~\footnote{\url{https://zenodo.org/record/7240482}} that are collected from three popular code hosting platforms (i.e., Github, GitLab, and Bitbucket). We further highlight 7,899 apps that have accessed CI/CD services.
    
    \item We have summarized a list of popular CI/CD services and implemented a script to automatically check if any of those services have been leveraged by given open-source Android app projects.
    
    \item We conduct a large-scale empirical study of CI/CD adoption in open-source Android apps and reveal to the community various interesting findings.
\end{itemize}




\section{Experimental Setup}

In this work, we propose to answer the following two research questions to experimentally understand the adoption of CI/CD in open-source Android apps.

    

\textbf{RQ1: To what extent is CI/CD adopted by Android projects? }
    We use Python scripts to automatically explore the usages of CI/CD services in all the collected open-source Android apps. Our experimental study reveals that (1) the majority of open-source Android apps do not involve CI/CD and (2) the apps that do adopt CI/CD services are mainly hosted on Github rather than GitLab or Bitbucket.
    
\textbf{RQ2: How are CI/CD services leveraged by Android app developers? }
    We go one step deeper to investigate the usage of CI/CD services by additionally looking at the building history of each Android app project.
    Our empirical investigation eventually reveals (1) a small number of apps have involved multiple CI/CD services and (2) nearly half of the apps adopted CI/CD services have not really used CI/CD services (i.e., no building history can be located).

\subsection{Dataset Preparation}

To support the experiments in answering these two research questions, we first prepare a dataset of open-source Android apps by mining the popular online code hosting sites.
Then, to help readers better understand this work, we provide more details about CI/CD and summarize the popular CI/CD services utilized by software developers in their daily development.
After that, we briefly introduce our methodologies for checking if a CI/CD service has been accessed by given Android apps.
To identify open-source Android projects we use the most famous open-source project hosting sites, namely Github, GitLab, and Bitbucket:

\textbf{Github.}
As is reported, Github~\cite{githubhome} has attracted more than 83 million developers and has held more than 200 million repositories~\cite{githubwiki} (at least 35 million of them are public repositories). It is non-trivial to iterate every public repository to determine if it is Android project. To help researchers and developers discover similar repositories, Github introduced topics~\cite{topicsintro} in 2017. Topics are labels that establish connections between Github repositories facilitating similar projects locating by type, technology, programming language, etc. We, therefore, utilize topic Android first to filter out non-Android repositories. However, there are still more than 90,000 repositories under topic Android and projects with the topic Android are not necessarily authentic Android repositories. They may be the Android design materials, books etc. To this end, we inspected every repositories containing topic Android to check if they have the Android special configuration file AndroidManifest.xml via the provided REST APIs~\cite{githubapi}. We, therefore, conclude Android repository if it contains the topic Android and has the Android specific configuration file AndroidManifest.xml~\cite{liu2020androzooopen} as is shown in the \textit{Android Repositories Identification} section of the Figure~\ref{fig:overview}. 

\textbf{GitLab.} GitLab~\cite{wikiGitLab,githubhome} is a DevOps platform provided by the open-core company GitLab Inc. It provides the core-functionalities under the MIT open-source license for public users while additional functionalities, such as vulnerability management, advanced security testing etc., are provided under proprietary license. It enables public developers to develop, secure, operate, and utilize CI/CD to complete their artefacts development. To harvest public Android repositories hosted on GitLab, we first listed all of the public repositories and then inspected if the project has the file named AndroidManifest.xml via the REST APIs provided by GitLab~\cite{GitLabapi}. 

\textbf{Bitbucket.} Bitbucket~\cite{bitbucketwiki,bitbuckethome} is also a Git-based code hosting service held by Atlassian. It provides commercial and free accounts both with unrestricted number of private repositories support. It is comprised of Bitbucket Cloud and Bitbucket Server. Bitbucket Server is the commercial software product and can be deployed on-premise with the license. Bitbucket Cloud is provided via URLs on Atlassian's Server. It gives the ability to complete the software development and testing via the exclusive CI/CD service for free accounts. Since we focus on Bitbucket Cloud, we would refer Bitbucket Cloud as Bitbucket for simplicity. To collect Android repositories on Github, we also listed all of the open-source projects hosted on Bitbucket, retrieved the latest commit for every project, listed all of the files for the latest commit, and finally inspected if the file of AndroidManifest.xml is provided or not via the self-hosted REST APIs~\cite{bitbucketapi}. 

In total, we have identified 84,475 open-source Android apps, including 78,245 Github apps, 341 GitLab apps, and 5,889 Bitbucket apps.

\subsection{CI/CD}
\begin{figure}[t!]
    \centering
    \includegraphics[width=\linewidth]{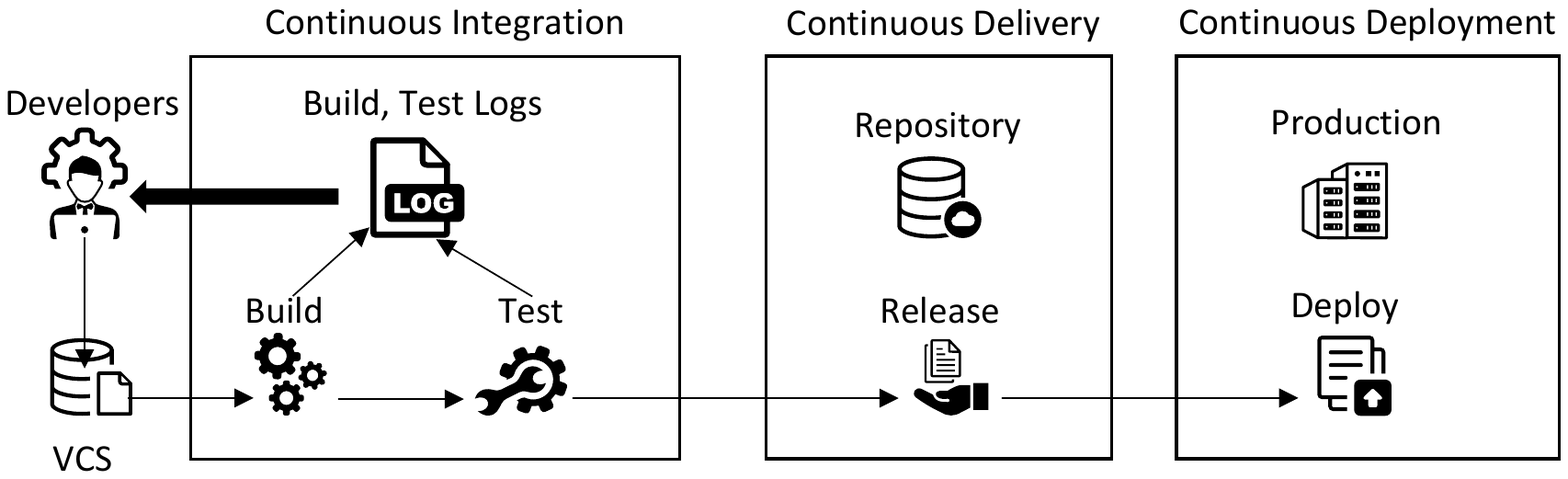}
    \caption{The typical working process for CI/CD systems.}
    \label{fig:cicdworking}
\end{figure}

Generally speaking, CI/CD accelerates the whole life cycle of artifact development via automatically building, testing, packaging, releasing, and deploying. Figure~\ref{fig:cicdworking} illustrates the typical working process of the CI/CD system. Modern software artifacts are expected to be maintained through Version Control Systems such as Git. Once new changes to the source code have been made by the developers and the certain trigger condition has been satisfied, the CI system will be involved to build and test the whole project automatically to check if such changes will induce any potential bugs. If there are build or test errors, developers will be notified with the build and test logs, which will help developers to locate the root causes to fix. Continuous Delivery/Deployment are tightly related concepts. They can be used interchangeably. Continuous Delivery usually means that the projects will be automatically uploaded to a repository (e.g., Github) and ready to deploy if they are built and tested successfully in CI. In comparison, Continuous Deployment refers to automatically deploying the released version to the production environment for customers.

CI/CD systems usually fulfill the automation by specifying specific jobs in the YAML-based configuration text files. YAML refers to YAML Ain't Markup Language~\cite{yamlofficial,yamlwiki}. It is a user-friendly data serialization language. Listing~\ref{lst:build_example} shows a build example excerpted  from~\texttt{PeopleAndService/BBasDriver-Android} ~\footnote{\url{https://github.com/PeopleAndService/BBasDriver-Android/blob/develop/.github/workflows/androidci.yml}}. To build the projects on Ubuntu-latest OS step by step, it first specify the running platform to Ubuntu-latest with the key word~\texttt{runs-on}. In the following steps, it checks out the source code, specify the version of Java, grant the execution permission for the Gradle wrapper (i.e., gradlew), and finally build the project. 

\begin{lstlisting}[caption={Github Action Build Example. Script is written via YAML.},label={lst:build_example},language=java]
build:
 name: Build Project
 runs-on: ubuntu-latest

steps:
 - uses: actions/checkout@v2
 - name: set up JDK 11
   uses: actions/setup-java@v2
   with:
    java-version: '11'
    distribution: 'adopt'
    cache: gradle

 - name: Grant execute permission for gradlew
   run: chmod +x gradlew
      
 - name: Build with Gradle
   run: ./gradlew build
\end{lstlisting}

\textbf{Existing Popular CI/CD Services.}
In this work, we perform an exploratory study (i.e., by searching online and reviewing the literature) to identify popular CI/CD services that are available for software developers to adopt.
Table~\ref{tab:list_cicd} summarizes our exploratory findings.
In total, we have found fifteen popular CI/CD services or tools, including three self-hosted services by our data source sites (cf. the last three rows).
The second column presents the proprietary information about the CI/CD services' configuration files (or directories where the configuration files are stored in).
This information could be leveraged to check whether the corresponding CI/CD service is adopted or not, which has actually been used in this work (details will be given in the next subsection).

\begin{table}[]
\footnotesize
    \centering
    \caption{List of popular CI/CD services and their specific configuration files.}
    \resizebox{\linewidth}{!}{
    \begin{tabular}{r | l}
    \hline
        Name              &  Configuration File  \\
        \hline
        Jenkins           &  *.yaml (content contains:~\texttt{jenkins:})   \\
        CircleCI          &   .circleci/config.yml  \\
        TeamCity          &   .BuildServer/teamcity-startup.properties   \\
        Bamboo            &   bamboo-specs/bamboo.yml  \\
        TravisCI          &   travis.yml             \\
        Codeship          &   codeship-services.yml \\
        GoCD              &   .gocd.yaml \\
        Wercker           &   wercker.yml   \\
        Semaphore         &   semaphore.yml    \\
        Nevercode         &   codemagic.yaml    \\
        Spinnaker         &   ./.hal-staging/   \\
        Buildbot          &   master.config   \\
        Github Actions    &   .github/workflow/xxx.yml    \\
        GitLab Pipelines  &   .gitLab-ci.yml     \\
        Bitbucket Pipelines &  bitbucket-pipelines.yml  \\
        \hline
    \end{tabular}}
    \label{tab:list_cicd}
\end{table}

\subsection{Methodology}

\begin{figure}[ht!]
    \centering
    \includegraphics[width=\linewidth]{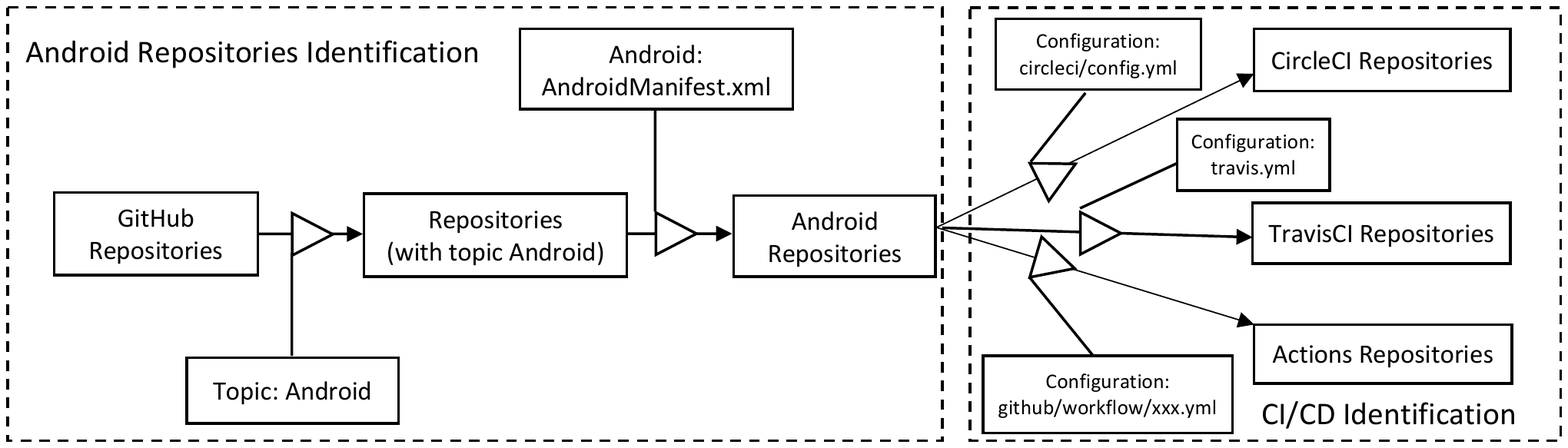}
    \caption{The working process of our approach -- using Github as an example for harvesting open-source Android app repositories and TravisCI, CircleCI and Github Actions as examples for locating CI/CD services.}
    \label{fig:overview}
\end{figure}

To determine if the Android repository was built by any of the popular CI/CD tools or services, we use Python scripts to traverse the file directories for each collected Android app project. Figure~\ref{fig:overview} represents the working process of our methodology (for selected CI/CD services only), for which we first leverage the file \emph{AndroidManifest.xml} (the essential configuration file for Android apps) to select Android app projects.
Then, we scan the repository files again to locate dedicated configuration files (as indicated in the second column in Table~\ref{tab:list_cicd}) to determine the usage of CI/CD services.
For example, we will regard a given app has leveraged TravisCI as long as we are able to locate the \emph{travis.yml} file.

\section{Results}
\label{sec:results}

\subsection{RQ1: \textit{To what extent is CI/CD adopted by Android projects?}}

In this research question, we explore what type of CI/CD services are practically utilized in open-source Android projects and then how many Android projects adopt different services. 
To this end, we launch our script to scan all the 84,475 open-source Android apps to inspect if the special configuration files summarized in Table~\ref{tab:list_cicd} exist or not.
The scanner eventually finds that 7,899 apps accessed CI/CD services, accounting for only 10.10\% (7,899/78,245) of Android app repositories.
This result suggests that Android developers still hesitate to integrate CI/CD services into their daily development.

\begin{table}[!ht]
    \centering
    \caption{The number of Android projects with CI/CD configurations.}
    \resizebox{\linewidth}{!}{
    \begin{tabular}{c|c|c|c|c|c}
        \hline
        \makecell{Data\\Source} & \textsc{TravisCI}  &   \textsc{CircleCI}  &  \makecell{Github\\\textsc{Actions}}  &  \makecell{Bitbucket\\\textsc{Pipelines}} & 
        \textsc{Total} \\
        \hline
        Github      &        3,153       &       927           &      4,095         &     0           &  7,870   \\  
        \hline
        Bitbuket    &         24         &        1            &        0           &     4            &    29     \\
        \hline
    \end{tabular}}
    \label{tab:ciusage}
\end{table}


\begin{figure}[!ht]
    \centering
    \includegraphics[width=0.8\linewidth]{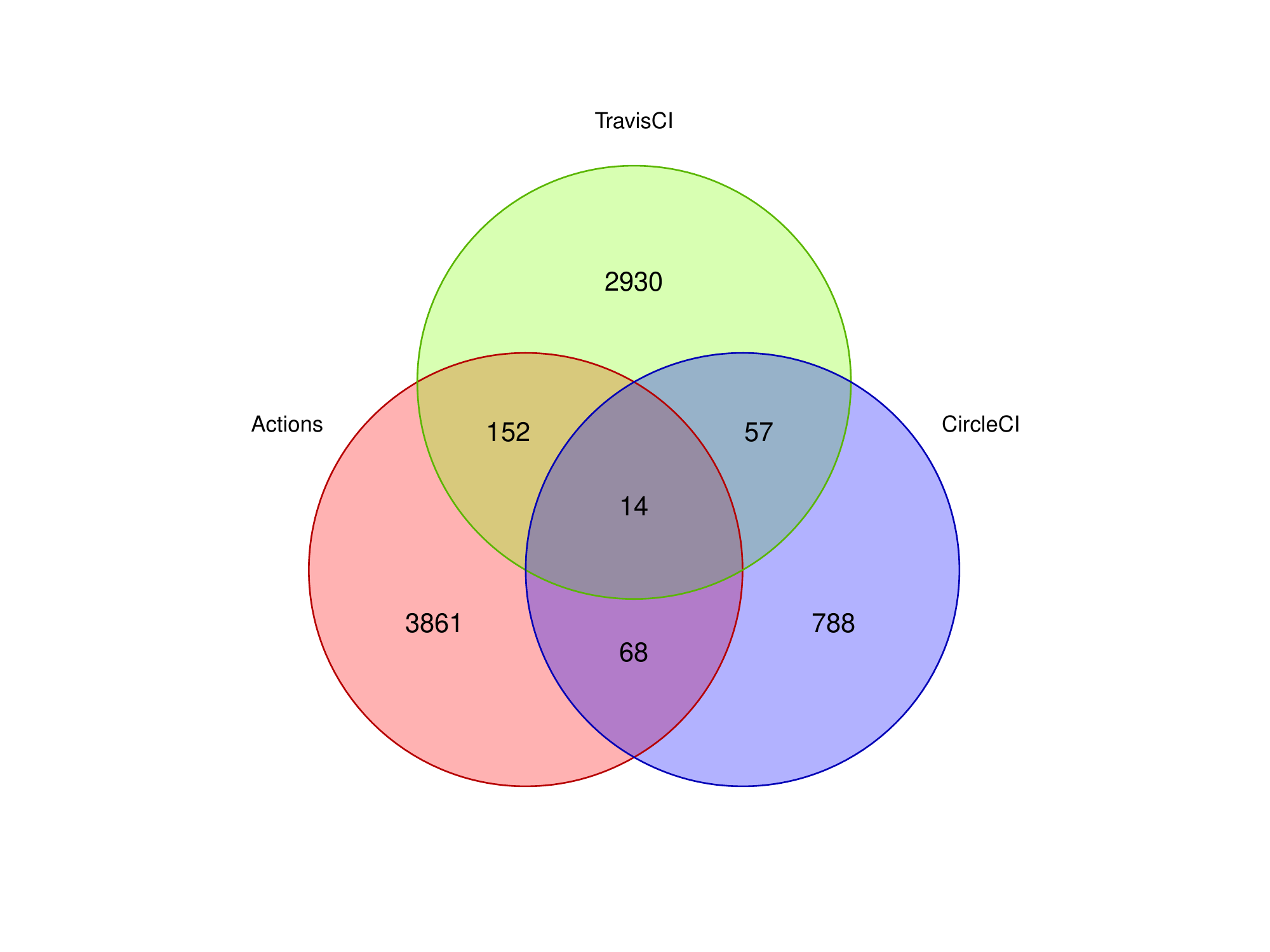}
    \caption{The distribution of the number of Android projects with three different CI/CD configuration files.}
    \label{fig:ciintersect}
\end{figure}


More interestingly, among the 15 selected popular CI/CD services, only four of them are adopted by open-source Android apps.
The four services are TravisCI, CircleCI, Github Actions, and Bitbucket Pipelines.
Table~\ref{tab:ciusage} further summarizes the actual number of Android app projects using these four CI/CD services w.r.t. the data sources. Different services are listed if the corresponding special files can be detected by our script in the Android projects. 
Specifically, the total number of Android projects with CI service from Github is 7,870, accounting for 10.06\% (7,870/78,245) of total Android apps.
For Bitbucket, there are only 29 apps (less than 0.5\% (29/5,889) of total apps) that have leveraged CI/CD services. 
Surprisingly, none of the Android projects on GitLab adopts any CI/CD services.
While for such projects that do access CI/CD services, Github~\textsc{Actions} is the most popular service that has been leveraged by 4,095 apps.
The second most popular service is \textsc{TravisCI}, which has been used by 3,177 apps.

\begin{figure}[!ht]
    \centering
    \includegraphics[width=0.8\linewidth]{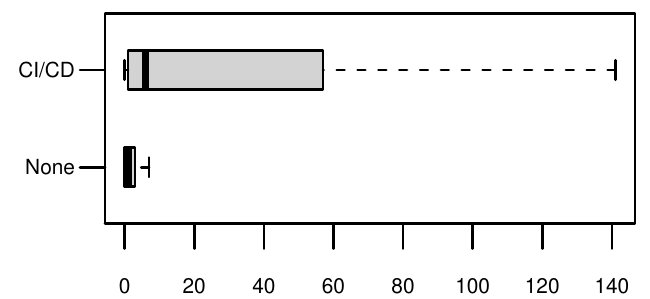}
    \caption{Number of stars given for Android projects with CI/CD service and without.}
    \label{fig:repo_withoutci}
\end{figure}

We further check if the adoption of CI/CD services is helpful in promoting the Android apps.
We achieve this by investigating whether apps with the help of CI/CD services will receive more stars on Github than those apps without accessing CI/CD services. 
Figure~\ref{fig:repo_withoutci} summarizes the experimental results. It illustrates the distribution of the number of stars given to Android projects configured with CI/CD services and those without any CI service. On average, the projects adopting CI/CD services receive more stars than those without any CI/CD service.
This difference has further been confirmed to be significant (at a significance level of 0.001) with an MWW\footnote{http://www.r-tutor.com/elementary-statistics/non-parametric-methods/mann-whitney-wilcoxon-test} test.
This result strongly suggests that CI/CD services are useful for Android app development, i.e., the corresponding apps are more popular and will be acknowledged by more developers.

\setlength{\fboxsep}{10pt}
\setlength{\fboxrule}{2pt}
\fcolorbox{gray!60}{gray!20}{%
  \parbox{0.90\linewidth}{%
  \textbf{RQ1 Findings:} \\
     Among 84,475 open-source Android apps collected from Github, Gitlab, and Bitbucket, only 7,899 of them (7,870 Github and 29 Bitbucket apps) have adopted CI/CD services, showing that CI/CD adoption in Android apps is still rare.
Furthermore, the fact that apps with CI/CD services generally receive more stars on Github than those without suggests that CI/CD is a useful technique for app developers to implement and maintain apps.}}


\subsection{RQ2: \textit{How are CI/CD services leveraged by Android app developers?}}


In this research question, we investigate the usage of CI/CD services during the development of Android projects. Observant readers may have noticed that the total number of Android projects (e.g. 7,870) is not equal to the sum of projects with different services (i.e. $3,153 + 927 + 4,095 = 8,175$).
This evidence suggests that some of the Android projects may contain more than one CI/CD service (with YAML-based configuration file). 
We confirm this by counting the number of CI/CD services (i.e., checking the number of YAML-based configuration files). 
Figure~\ref{fig:ciintersect} summarizes the distribution of the number of Android projects that involve three different types (i.e., TravisCI, CircleCI and Actions) of CI/CD configuration files. Indeed, for Github Android projects, there are 291 ones containing more than one special configuration file (with a rate of 3.56\% (291/8,175)). 




Given the fact that the adoption of CI/CD services in the source code does not guarantee that such services would be deployed in the process of Android development. The aforementioned finding motivates us to investigate the exact usage of CI/CD services in practice. To this end, we retrieve the build history provided by the corresponding CI/CD service. We consider the project was truly built with CI/CD services if we could successfully retrieve its build log (cf. Figure~\ref{fig:cicdworking}).
Table~\ref{tab:ciutilize} summarizes the experimental results. In this work, we only consider the Android projects hosted on Github\footnote{We exclude Android projects from GitLab and Bitbucket as the number of such projects is limited.} and those were successfully built at least once with CI services (i.e., build logs can indeed be located). 
As shown in the table, \emph{Actions} occupies the maximum proportion, with a rate of 96.63\% (3,957/4,095). Except for the usage of \emph{Actions}, the percentage of Android projects adopted \emph{TravisCI} and \emph{CircleCI} are 20.01\% (631/3,153) and 16.28\% (151/927), respectively. 
This result suggests that compared with \emph{TravisCI} and \emph{CircleCI}, Github \emph{Actions} either is easier to be successfully configured or receives better support from Github.
Hence, we suggest that Android app developers adopt Github \emph{Actions} as the CI/CD service if they decide to host their apps on Github.

\begin{figure}[!ht]
    \centering
    \includegraphics[width=0.8\linewidth]{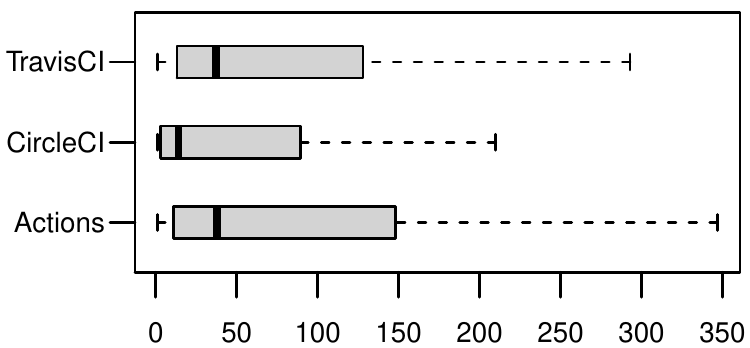}
    \caption{Number of executions of different CI/CD services.}
    \label{fig:cileverage}
\end{figure}

Furthermore, we investigate the CI/CD processes from the retrieved build history for those apps that have successfully run CI/CD services.
Figure~\ref{fig:cileverage} illustrates the experimental results. It shows the execution counts of the configured workflows fulfilling the automated process among three CI/CD services. The median values of the execution are 38, 14, and 37 for Github Actions, CircleCI, and TravisCI, respectively. Compared to the recent studies~\cite{zhang2019large,zhang2022buildsheriff,gallaba2022lessons,gallaba2018noise} in services CircleCI and TravisCI, developers in Android community may not fully exploit the ability as 
the execution counts are relatively small.

\begin{figure}[!ht]
    \centering
    \includegraphics[width=0.8\linewidth]{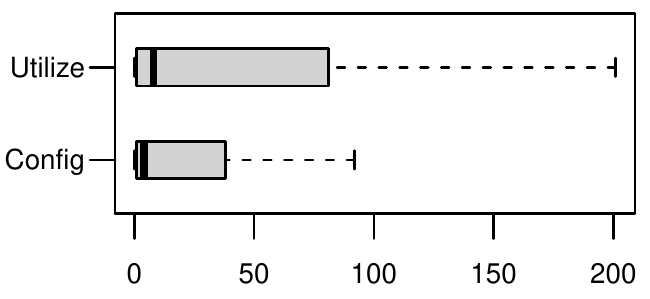}
    \vspace{3mm}
    \caption{Number of stars given for Android projects only with CI/CD configuration and those with CI/CD utilization.}
    \label{fig:repo_withoutexe}
\end{figure}

The fact that only around half of the apps (4,706) adopted CI/CD services have actually run them in practice motivates us to revisit the usefulness study of CI/CD services conducted in the previous section.
Specifically, we explore the difference of the number of stars given to the Android projects only configured with CI/CD special files and those processed with CI/CD services. Figure~\ref{fig:repo_withoutexe} shows the distribution of the number of stars for the projects only with CI/CD configuration and those processed with CI/CD service. The median value for the projects only with CI/CD configuration is four while for the projects handled with CI/CD service is eight. The experimental result confirms that the projects processed with CI/CD services indeed become more popular, i.e., gain more attention from other developers.

\begin{table}[t]
    \centering
    \caption{The number of Android projects with executions of CI/CD services}
    \resizebox{\linewidth}{!}{
    \footnotesize
    \begin{tabular}{c|c|c|c}
        \hline
        \textsc{TravisCI}  &   \textsc{CircleCI}  &  \textsc{Actions}  &  \textsc{Total} \\
        \hline
               631         &         151          &      3,957         &      4,706   \\  
        \hline
    \end{tabular}}
    \label{tab:ciutilize}
\end{table}


\setlength{\fboxsep}{10pt}
\setlength{\fboxrule}{2pt}
\fcolorbox{gray!60}{gray!20}{%
  \parbox{0.90\linewidth}{%
  \textbf{RQ2 Findings:} \\
     Only 59.60\% (4,708/7,899) Android projects with a CI/CD service actually executed at least once the adopted services. Among these repositories,  developers tend to use Github~\textsc{Actions} to fulfill their automation process of building, testing, and deploying etc. 
Compared to projects with CI/CD configuration, the projects that actually executed CI/CD services  gain more attention from other fellow developers.}}

\section{Threats to Validity}
\label{discussion_threat}

The primary threat to the validity concerns the approach to collecting the Android projects. To curate the Android dataset, we only consider the projects with the topic Android and count them as Android if it contains the specific configuration file~\texttt{AndroidManifest\\.xml}. However, we would miss some Android projects if they do not provide the topic Android.
Second, our study is solely based on the files recorded on the online code hosting sites. It is highly possible that some CI/CD services are only adopted in developers' local environments. Such cases are directly ignored in this work. Nonetheless, we believe such cases are rare in practice (as CI/CD often needs to work with version control systems such as git), and hence our experimental results should still be representative.
Furthermore, we can not guarantee that the open-source Android apps are all under good maintenance and the conclusion that the projects with CI/CD services imply more popular can be drawn under any circumstances. Nevertheless, our empirical findings should still be valid as long as the number of apps out of maintenance is within a small portion in our dataset. We plan to explore the impact caused by the apps out of maintenance and other factors influencing the relationship between the number of stars and the adoption of CI/CD services further in our future work.



\section{Related Work}
\label{sec:relatedwork}

Travis-CI, one of the most popular CI platforms for open-source projects, has attracted many different researchers. 
Beller et al.~\cite{beller2017travistorrent} provides to the research community a dataset called~\textsc{TravisTorrent}, which contains 2,640,825 Travis builds from more than 1,000 open-source Github projects. 
Going one step further, Gallaba et al.~\cite{gallaba2018noise} utilized the~\textsc{TravisTorrent} dataset and extracted additional data for the projects in the dataset from the REST APIs provided by~\textsc{Travis-CI}. 
They found that the build results are noisy and heterogeneous. 
Widder et al.~\cite{widder2019conceptual} conducted a literature review of 37 papers focusing on CI and replicated the experiments in the papers. In addition, they surveyed 12 interviews and received 132 responses. They summarized eight factors that resulted in abandoning or switching the~\textsc{Travis CI} among 6,239 Github repositories and proposed potential research directions for researchers and possible improvements for CI providers. 
Zhang et at.~\cite{zhang2019large} conducted a large-scale empirical study and concluded that CI build failures caused by compiler errors are common even in successful builds. 
They~\cite{zhang2022buildsheriff} further presented a large-scale study to triage test failures among CI Github Java projects and proposed an approach~\textsc{BuildSheriff} to automate such classification. 

Hung et al.~\cite{hung2019continuous} and Giang et al.~\cite{giang2020automated} discussed the automation of Android Application build and test and presented the implementation of the automation script for building and testing via CI tool~\textsc{CircleCI}. The example of the script implementation would boost the Android app development by providing guidance for the developers and automating the build and test per se.
Gallaba et al.~\cite{gallaba2022lessons} conducted an empirical study on 22.2 million builds over 7,795 open-source projects adopted~\textsc{CircleCI} from 2012 to 2020. Their quantitative analyses on the job build concluded that the service consumers spent most of their time working on specific build processes (e.g., dependency retrieval), and most of the build time was spent on compilation and testing. Their findings suggest that service consumers would benefit most if the long-duration builds, compilation build, and testing build could be accelerated. The robustness of the CI service would also be improved if the misconfiguration of the build scripts and service availability issues could be handled properly.

Hilton et al.~\cite{hilton2016usage} leveraged three complementary methods to investigate the usage of CI services in Github open-source projects. They curated a dataset containing 34,544 open-source projects from Github and found that most of the projects adopted~\textsc{TravisCI}. Besides~\textsc{TravisCI}, projects developers also utilize~\textsc{CircleCI},~\textsc{AppVeyor},~\textsc{ClouldBees}, and~\textsc{Werker} among the collected dataset. Moreover, their analyses revealed that project developers did not adopt CI services because they were unfamiliar with any of the CI services, and the projects with CI services were released twice as often as those without. In addition, they provided suggestions to researchers, developers, and CI service providers.

\section{Conclusion} 
\label{sec:conclusion}
In this study, we conducted an extensive empirical investigation of the utilization of CI/CD services in open-source Android repositories from popular code hosting sites, including Github, GitLab, and Bitbucket. To the best of our knowledge, we are the first to focus on use of CI/CD services among Android app projects. Our study reveals that most app developers do not adopt CI/CD services in their development. Among the Android projects handled by the CI/CD services, the most popular three services are Github~\textsc{Actions}, \textsc{TravisCI}, and \textsc{CircleCI}. Interestingly, projects with CI/CD services are generally more popular than those projects that do not involve CI/CD. Such projects are even more popular if they further guarantee that the CI/CD processes are well executed compared to those that do not.

\begin{acks}
This work is supported by ARC Laureate Fellowship FL190100035, ARC Discovery Early Career Researcher Award (DECRA) project DE200100016, and a Discovery project DP200100020.
\end{acks}

\balance
\bibliographystyle{ACM-Reference-Format}
\bibliography{main.bib}

\end{document}